\documentclass[12pt]{article}
\oddsidemargin = \evensidemargin
\usepackage{epsfig}
\usepackage{amssymb}
\usepackage{colordvi}

\begin{document}

\title{\bf Inherent Global Stabilization \\ of Unstable Local Behavior \\
in Coupled Map Lattices}

\author{Harald Atmanspacher$^{1,2}$ and Herbert Scheingraber$^1$ \\ \\
1 Center for Interdisciplinary Plasma Science, \\
Max-Planck-Institut f\"ur extraterrest\-rische Physik, \\ 85740 Garching, Germany \smallskip \\
2 Institute for Frontier Areas of Psychology and Mental Health, \\ 
Wilhelmstr.~3a, 79098 Freiburg, Germany}

\date{}
\maketitle

\bigskip
\centerline{Accepted for publication in}
\centerline{{\it Intern.~Journal of Bifurcation and Chaos}}
\vskip 3cm

\begin{abstract}

The behavior of two-dimensional coupled map lattices is studied with respect to 
the global stabilization of unstable local fixed points without
external control. It is numerically shown under which circumstances
such inherent global stabilization can be achieved for  both
synchronous and asynchronous updating. Two necessary 
conditions for inherent global stabilization are derived analytically. 

\end{abstract}


\vfill\eject

\section{Introduction} 

Coupled map lattices (CMLs) are arrays of states whose value is continuous,
usually within the unit interval, over discrete space and time. Starting with
Turing's seminal work on morphogenesis [Turing 1952], they have
been used to study the behavior of complex spatiotemporal systems for
50 years. More recently, Kaneko and collaborators have established many
interesting results for CMLs (cf.~Kaneko [1993]) as generalizations of 
cellular automata, whose state values are discrete.    

One key motivation for modeling spatiotemporally extended systems with CMLs
is to simplify the standard approach in terms of partial differential
equations. And, of course, CMLs would not have become accessible without
the rapid development of fast computers with large storage capacities.
Within the last decades, CMLs have been applied to the study of areas as diverse
as social systems, ecosystems, neural networks, spin lattices, Josephson junctions,
multimode lasers, hydrodynamical turbulence, and others (cf.~the special 
journal issues {\it CHAOS} {\bf 2}(3), 1992, and {\it Physica D} {\bf 103}, 1997).       

A compact characterization of a CML over two
spatial dimensions with one time parameter is given by:
\begin{equation}
g_{n+1}(x_{ij}) = (1-\epsilon) f_n(x_{ij}) + {\epsilon\over N} \sum_{k=1}^N f_n(x_k)  
\end{equation}
For $f_n(x)$, the iterative logistic map $x_{n} = rx_{n-1}(1-x_{n-1})$ is mostly 
used, where $r$ is a control parameter, $0 <r \le 4$, and $n$ represents the time step
of the iteration.
The indices $i,j$ are used to label the position of each cell (or site) in the 
lattice. $N$ is the number of cells defining the neighborhood 
of each cell (with periodic boundary conditions),
and $k$ runs over all neighboring cells. The parameter $\epsilon$
specifies the coupling between each cell and its neighborhood (and is usually 
considered as constant over time and space). Thus, the value 
of $g_{n+1}(x_{ij})$ is a convex combination of the value at each individual cell 
and the mean value of its local neighborhood.   

For $\epsilon\rightarrow 0$, there is no coupling at all; hence, local
neighborhoods have no influence on the behavior of the CML.  This situation
represents the limiting case of $N_{tot}$ independently operating local objects 
at each lattice
site. In the general case $0< \epsilon <1$,  the independence of individual cells is 
lost and the lattice behavior is governed by both local and 
global influences. CMLs with a maximal neighborhood, $N \approx N_{tot}$, 
are often denoted as globally coupled maps.    
If the coupling is maximal, $\epsilon\rightarrow 1$,
the behavior of the entire CML is determined by global properties alone (mean field 
approach).
 
The second term on the rhs in Eq.~(1) contains the states of the neighboring 
map sites at the same time step $n$ at which the first term specifies the state 
of the site whose neighborhood is considered. This type of coupling, assuming
a vanishing transmission time $\Delta t\rightarrow 0$, is sometimes
called ``future coupling'' [Mehta \& Sinha 2000]
since it refers to a situation in which the neighborhood 
states are treated as if they act back from future to present. 
In order to take a finite transmission time $\Delta t > 0$ into account, 
one can modify the second term in Eq.~(1)
such that $f_n(x_k)$ is replaced by $f_{n-1}(x_k) = x_k$. In this way, past states 
in the neighborhood of a site are considered to act on the present state
of a given site with limited signal speed so that interactions are delayed
rather than instantaneous. Corresponding coupling scenarios, which have 
recently been studied by Mehta \& Sinha [2000], Masoller et al.~[2003], Li et al.~[2004], 
and Atay et al.~[2004], will be focused at in the present paper.    

Another time scale important for the physical interpretation of Eq.~(1) is
the time interval $\Delta \tau$ assumed for the updating mechanism, i.e.~for
the physical integration of signals from the neighborhood states with the
state considered. If signals between cells are
transmitted much slower than the time scale assumed for the updating mechanism,
$\Delta \tau \ll \Delta t$, the updating can be implemented (almost) instantaneously,
or synchronously. If this is not the case, $\Delta \tau \gtrsim \Delta t$, 
updating must be implemented in an asynchronous way. This entails the 
additional problem of determining a proper updating sequence, which can be 
random or depend on particular features of the situation considered.

Most of the work on CMLs 
published in this respect (cf.~Kaneko and Tsuda [2000]) was based on synchronous 
updating. For asynchronous updating as, for instance, studied by Lumer \& Nicolis 
[1994], it was found that the behavior of CMLs differs strongly from that of CMLs with 
synchronous
updating. Additional results for asynchronous updating were reported by Marcq et 
al.~[1997], Rolf et al.~[1998], and Mehta \& Sinha [2000].
Asynchronous updating rules have been suggested as particularly relevant for 
neurobiological networks.
The relevance of CMLs with functions with quadratic maximum (such as the
logistic map) as models for neurobiological networks was recently substantiated
by novel results concerning a non-monotonic (rather than sigmoid) transfer function 
for individual neurons (Kuhn et al.~[2004]). 

As a common feature of the (so far) few studies of asynchronous updating, it has 
been reported that it facilitates the synchronization and stabilization of CMLs
decisively. In particular, Mehta \& Sinha [2000] demonstrated that the dynamics 
at individual lattice cells is strongly synchronized by coupling among cells.
In this contribution we will focus on specific stability properties of CMLs rather 
than their synchronization and evolving patterns. In particular, we will show that 
unstable fixed points at individual cells can be stabilized 
as a consequence of their coupling to neighboring unstable fixed points.     
(For more general issues concerning the asymptotic stability of densities
in coupled map lattices compare Mackey \& Milton [1995].) 

Such a stabilization is of particular interest since it is independent of external
control mechanisms. The global stabilization of unstable local behavior operates 
inherently, without external adjustment, 
once the coupling is strong enough. At a speculative level, such a possibility
was indicated by Atmanspacher \& Wiedenmann [1999]. It represents a
powerful alternative to external control procedures in the style of ``controlling
chaos'' (Ott et al.~[1992]). 

Another motivation to study globally stabilized local instabilities derives from
work on the perception of bistable stimuli in cognitive contexts. Recently,
a specific neural correlate in event-related potentials was discovered for
the switching process between the two different mental representations of a 
bistable stimulus (Kornmeier et al.~[2004]). As outlined by Atmanspacher [1992],
the unstable state between the two (meta-)stable representations might be
an interesting candidate for particular stabilization procedures.             
 
The following section 2 presents the results of numerical 
simulations of CMLs with synchronous and asynchronous updating, for different
coupling strengths $0<\epsilon<1$, and for different types and sizes of 
neighborhoods. It will be demonstrated that stabilizing effects for unstable 
local behavior are generic for asynchronous updating with strong enough 
coupling. For synchronous updating, it depends on the
type of neighborhood whether and when such a stabilizing effect occurs.

In section 3, the stabilizing effect for local unstable behavior
is considered from an analytical point
of view. It will be shown how the inherent global stabilization of  
unstable local fixed points can be understood in terms of a  
``squeeze-and-shift'' procedure applied to the logistic map. Two necessary 
conditions for the stabilization will be derived and 
significant parameters will be discussed.    

Section 4 summarizes and concludes the paper, and some perspectives will 
be addressed.

\section{Numerical Results}

In this section, we present results from numerical simulations of two-dimensional 
coupled map lattices according to Eq.~(1). Since the focus of this contribution
is on the stabilization of unstable behavior, we have to work within a parameter
range in which the behavior of the logistic maps at each lattice site is unstable.        

The logistic map $x_{n} = rx_{n-1}(1-x_{n-1})$ 
has two critical points, one at 0 and one at ${r-1\over r}$. 
The stability properties of these critical points are directly related 
to the derivative of the function $x \mapsto rx(1-x)$ at each of them. If the absolute value
of the derivative
is smaller (greater) than 1, then the critical point is stable (unstable). 
Hence, the critical point at 0 is a stable fixed point for $r<1$ and unstable
for $r\ge 1$. The critical point at ${r-1\over r}$ is stable for $r<3$ and 
unstable for $r\ge 3$.   

For our investigations, 
we focus on the more interesting unstable fixed point
at ${r-1\over r}$ and use the control parameter $r=4$ to demonstrate the results. 
(As far as the topic of this contribution is concerned, there is no basic
difference in behavior for other values of $r$ as long as $r>3$.)
The corresponding unstable fixed point is located at 0.75.
We study the distribution of state values of a lattice of size $50\times 50$ 
($N_{tot}=2500$ cells with random initial conditions) after a number of iterations which
is large enough that transients have died out, usually after 10000 iteration steps. 

We consider different kinds of neighborhoods according to 
the second term of Eq.~(1). Results for both von Neumann neighborhoods and 
Moore neighborhoods will be presented, each of both order 1 and 2. A von
Neumann neighborhood of order 1 includes the $N=4$ vertically and horizontally
nearest neighbors of a given site. A Moore neighborhood of order 1 includes the
4 diagonal nearest neighbors in addition, hence covering a square of $N=8$ cells 
in total. A von Neumann
neighborhood of order 2 is constructed by a Moore neighborhood of order 1 plus
the vertical and horizontal second neighbors of a given site, hence it consists
of $N=12$ cells in total. A Moore neighborhood of order 2 covers, in addition, all cells
covering a square of side length 5, hence $N=24$ cells in total.     

As mentioned in the introduction, the behavior of CMLs depends on the way in which
the values at each cell are updated from one to the next iteration step. As two
basic types of updating, we distinguish between synchronous updating, where 
all values are calculated subsequently but updated at once, and asynchronous 
updating, where all values are updated in the sequence in which they are calculated.
For the latter procedure, it is crucial how the sequence is defined. 
In case of asynchronous updating, those cells which are 
already updated affect the behavior of the CML before the update providing the next
iteration step is complete. This does not happen in case of synchronous updating.  

\subsection{Synchronous Updating}

Figures 1 and 2 show histograms for the distribution of state values at individual cells
after 10000
iteration steps for a von Neumann neighborhood of order 1 and a Moore neighborhood 
of order 1, for coupling strengths $\epsilon = 0.5$ (Figs.~1a, 2a) and $\epsilon = 0.8$
(Figs.~1b, 2b). Figs.~1a and 2a show two differently pronounced peaks right and left of 
the unstable fixed point at 0.75, indicating oscillatory behavior of the overall state distribution
of the CML. (For neighborhoods of order 2, similar behavior is observed.)
The situation is different in Figs.~1b and 2b. While the oscillating behavior
is maintained for the von Neumann neighborhood of order 1, the Moore neighborhood of 
order 1 produces a stabilization of the CML at the unstable fixed point at 0.75 at each cell.
(Both neighborhoods provide stabilization at 0.75 if they are of order 2.) 

Fig.~3 gives an overview representation of this behavior, a so-called stability diagram,  
for the full range $0\le \epsilon \le 1$. The vertical axis indicates mean values of the state 
distribution left and right of the unstable fixed point at 0.75, respectively, which are 
averaged over ten different sets of random initial conditions for the CML. The four 
curves represent the four different types of neighborhood. 

The behavior shown in Figs.~1a,b and 2a,b corresponds to one point for each histogram
in the stability diagram Fig.~3. Error bars are of the size of the symbols. 
It is clearly visible that the von Neumann neighborhood of 
order 1 does not reach stabilization at 0.75 but stays bimodal until the mean value
of the distribution of initial states ($\approx$ 0.5) is obtained for $\epsilon = 1$ 
(this is the case for all types of neighborhood).
For all other types of neighborhood considered, there is a critical value of $\epsilon$
at which stabilization sets in. This critical value decreases for an increasing order of the 
neighborhood (even beyond 2, although this is not shown here). 
For small values of $\epsilon$, the behavior of all four stability curves is essentially 
identical.           

\renewcommand{\baselinestretch}{0.85}
\begin{figure}
\epsfig{figure=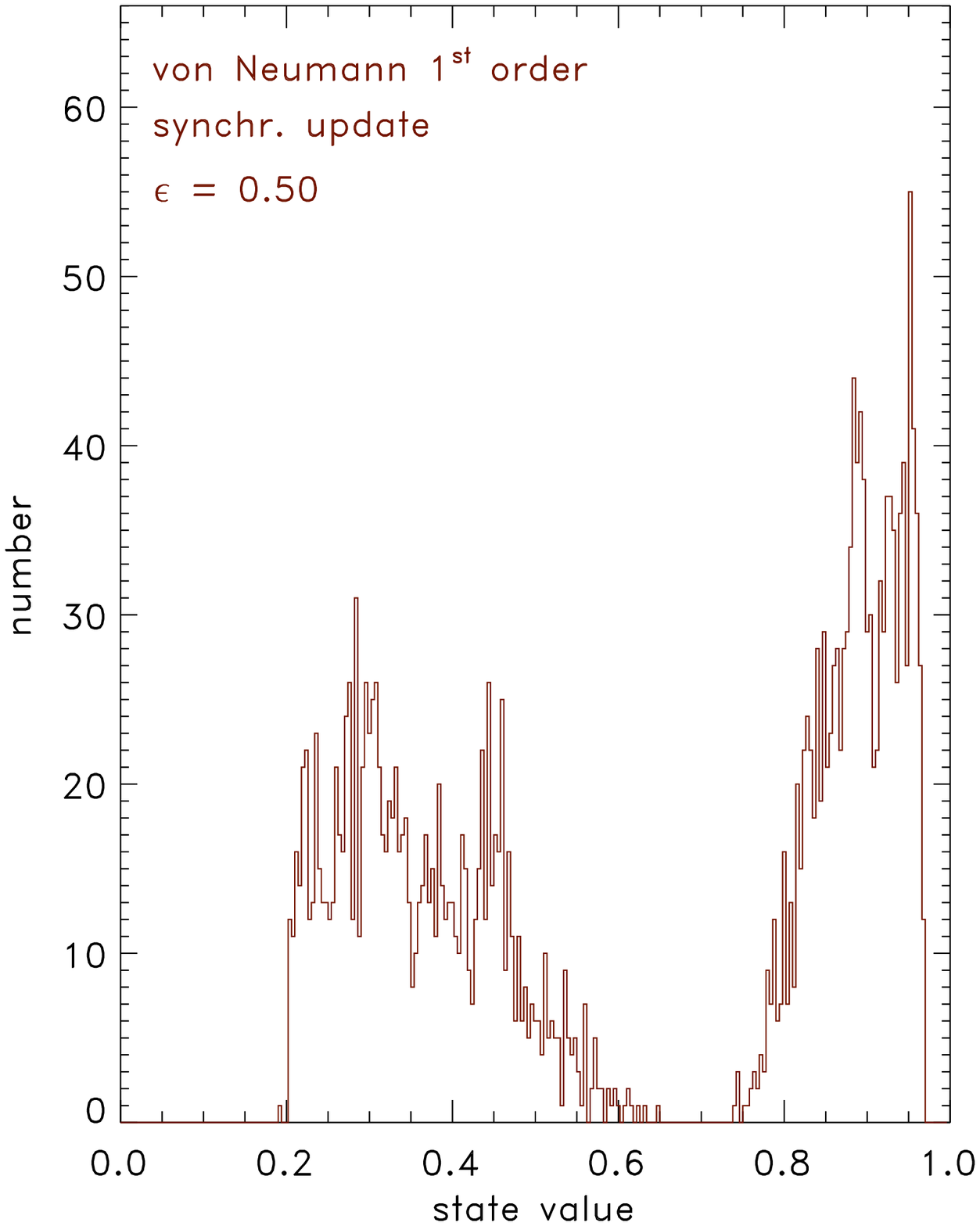,scale=0.45}  
\epsfig{figure=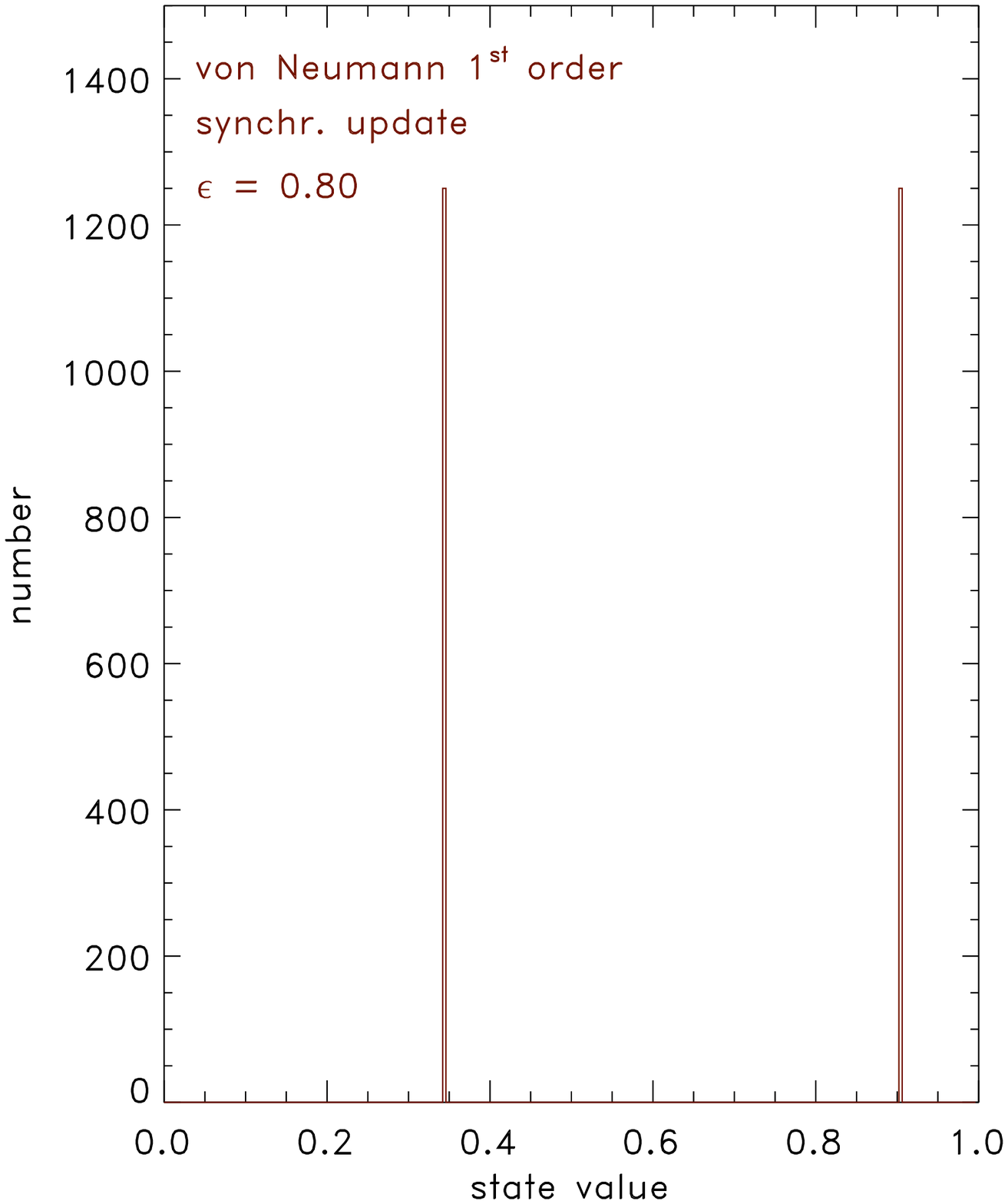,scale=0.45}  
\begin{quote}
{\footnotesize Figure 1: State histograms for synchronously updated CMLs with a von Neumann
neighborhood of order 1 for coupling strengths (a) $\epsilon = 0.5$ (left) and (b) $\epsilon = 0.8$ (right). 
The control parameter of the logistic map is set at $r=4$, and the number of iterations is 10000.}    
\end{quote}
\bigskip
\epsfig{figure=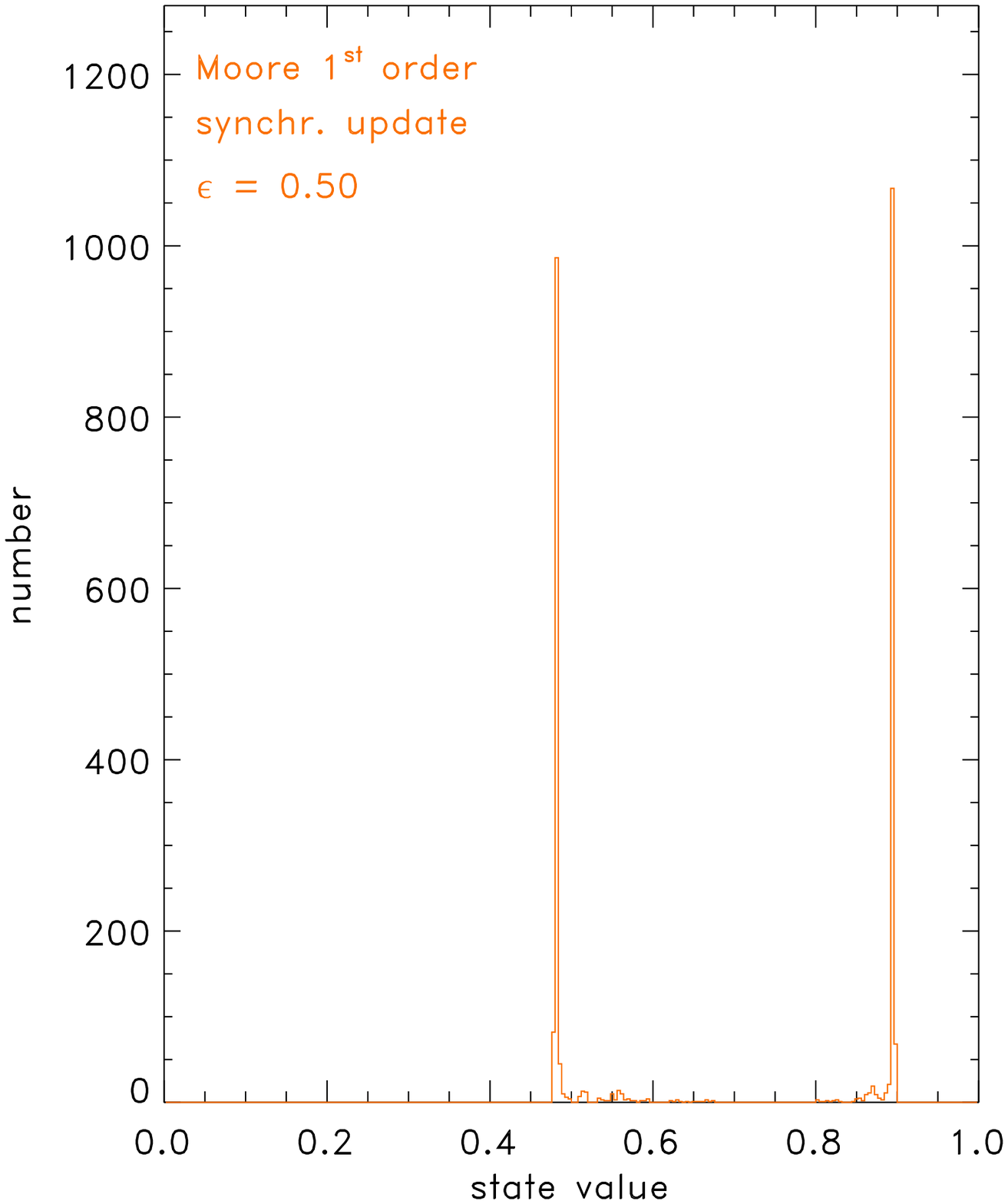,scale=0.45}  
\epsfig{figure=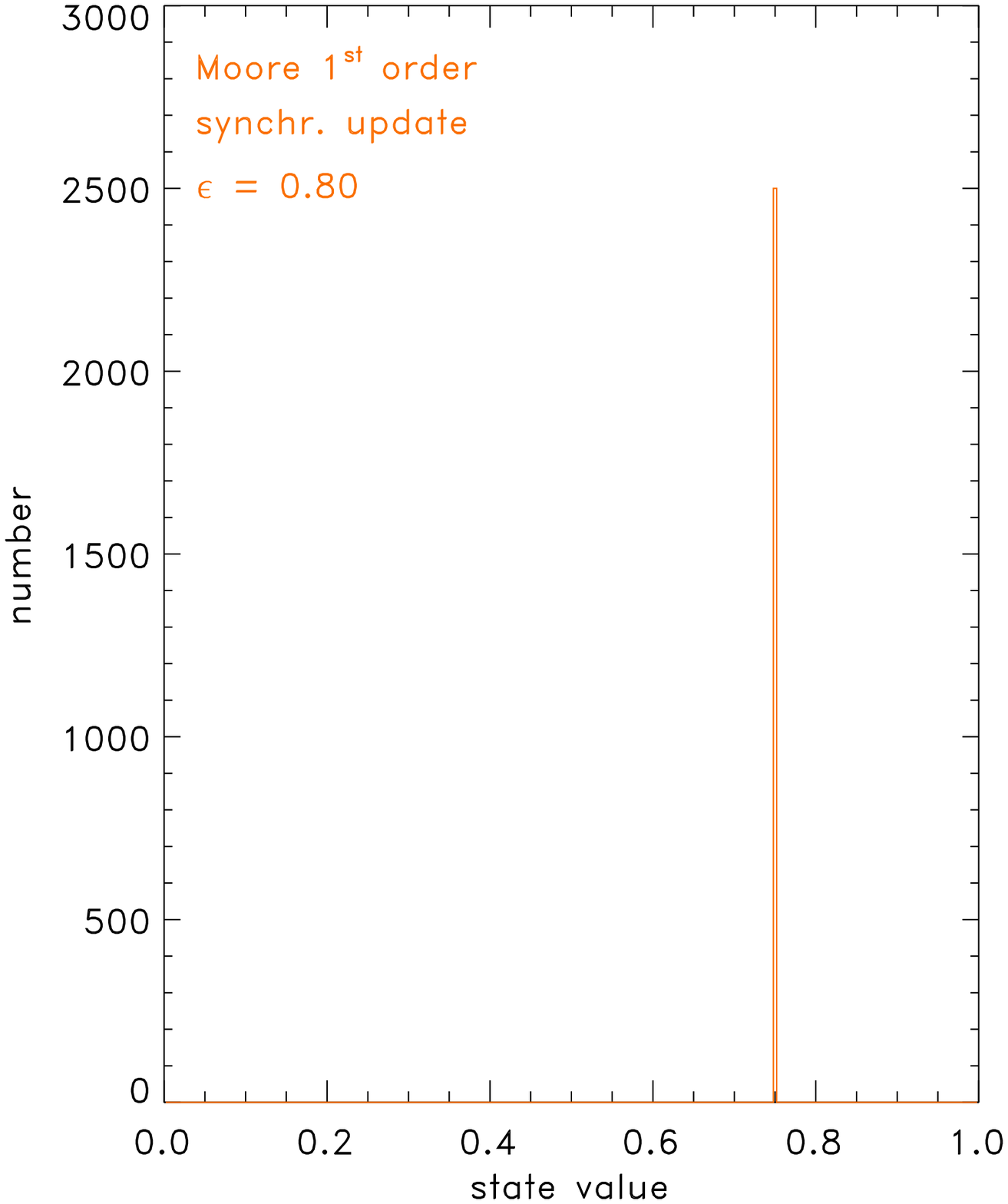,scale=0.45}  
\begin{quote}
{\footnotesize Figure 2: State histograms for synchronously updated CMLs with a Moore 
neighborhood of order 1 for coupling strengths (a) $\epsilon = 0.5$ (left) and (b) $\epsilon = 0.8$ (right). 
The control parameter of the logistic map is set at $r=4$, and the number of iterations is 10000.}    
\end{quote}
\end{figure}

\renewcommand{\baselinestretch}{0.85}
\begin{figure}[h]
\begin{center}
\epsfig{figure=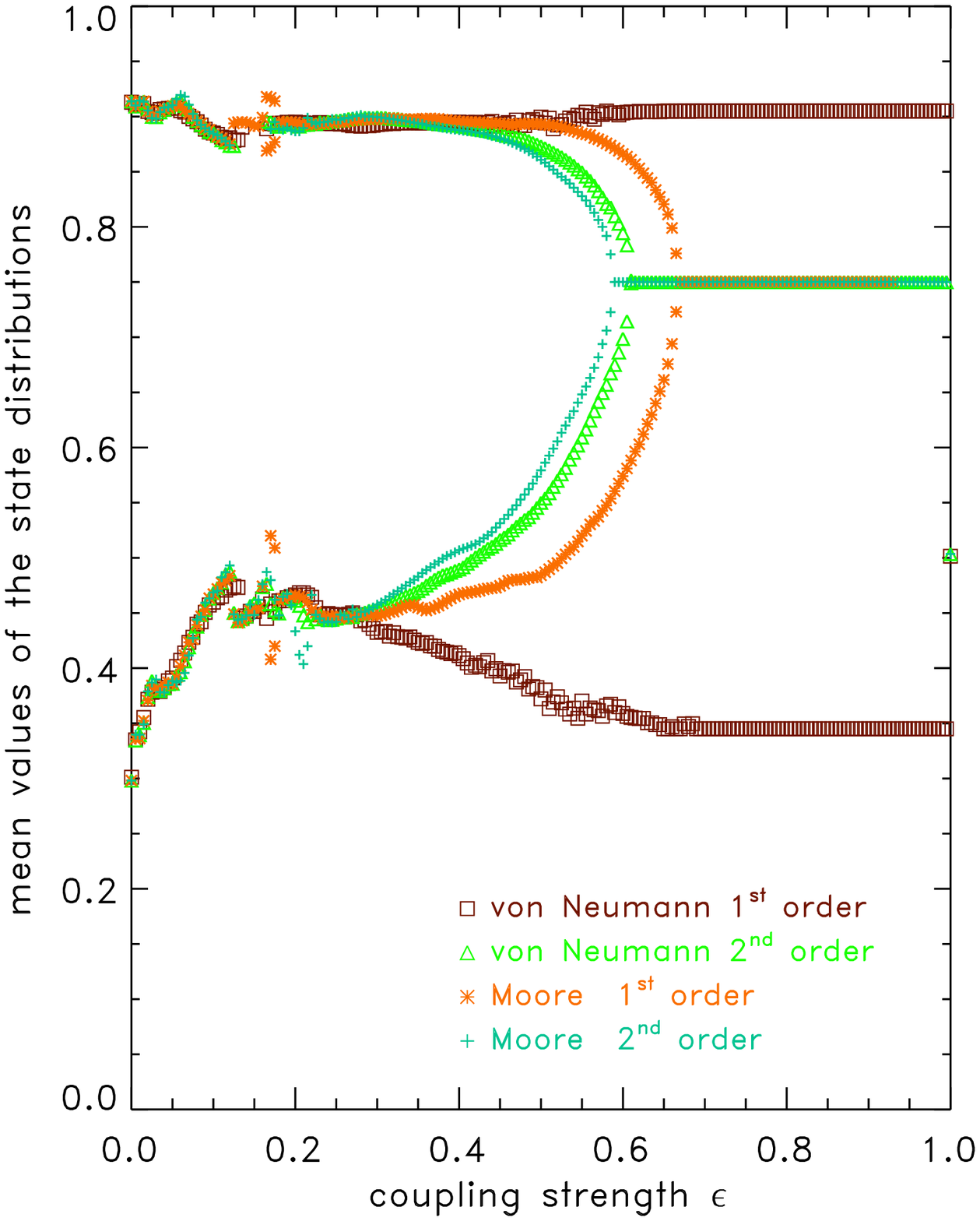,scale=0.6}  
\end{center}
\begin{quote}
{\footnotesize Figure 3: Stability diagram for synchronously updated CMLs with different 
types of neighborhoods. Mean values of the state distribution right and left of the unstable
fixed point at 0.75, averaged over ten sets of random initial conditions, are plotted
versus the coupling strength $\epsilon$. The control parameter of the logistic map is
set at $r=4$, and the number of iterations is 10000.}    
\end{quote}
\end{figure}

As a result, local unstable behavior of the logistic map (at the unstable fixed point at 0.75) 
can be stabilized due to the influence of neighborhoods in coupled map lattices under synchronous 
updating. Such stabilization is observed for both homogeneous and inhomogeneous
perturbations. The only exception found to this rule is a von Neumann neighborhood of 
order 1 which fails to provide stabilization. Considering higher order neighborhoods of different
kinds provides a substantial extension of 
the observations by Mehta \& Sinha [2000] whose studies of synchronous updating were
restricted to von Neumann neighborhoods of order 1.            
 
\subsection{Asynchronous Updating}

For an asynchronous updating procedure, the updating sequence is crucial and should
be adapted to the situation which is modeled by the implemented CML. In cases which can
be considered more or less spatially homogeneous, a random selection of the updating
sequence is naturally plausible. 
In the particular case of neurobiological applications, i.e.~neural networks, other 
sequencing mechanisms may be preferrable. For instance, it might be appropriate to use a 
value-dependent asynchronous update (similar to Hopfield [1982]), in which cells with higher 
values are updated first.  

\renewcommand{\baselinestretch}{0.85}
\begin{figure}
\epsfig{figure=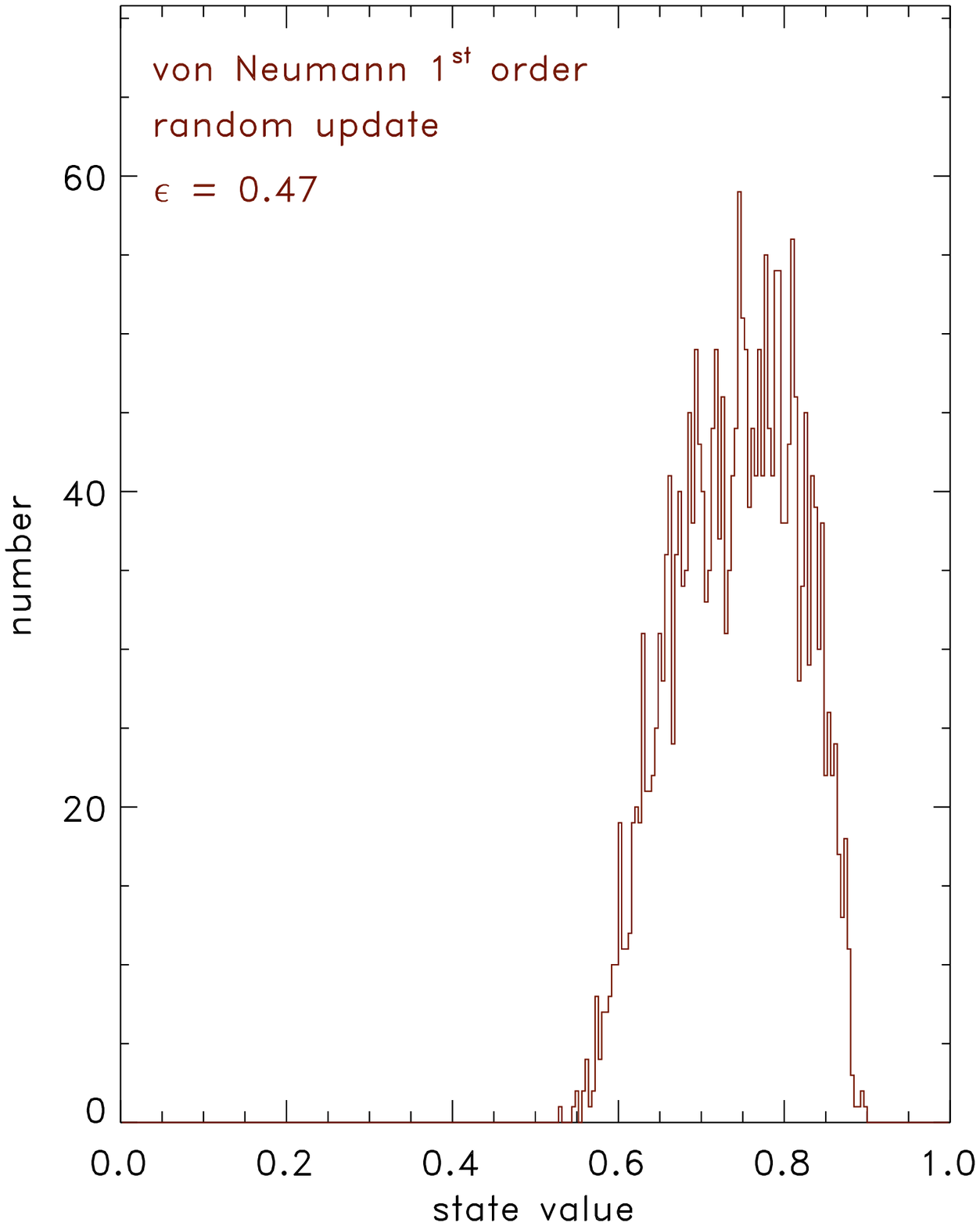,scale=0.45}  
\epsfig{figure=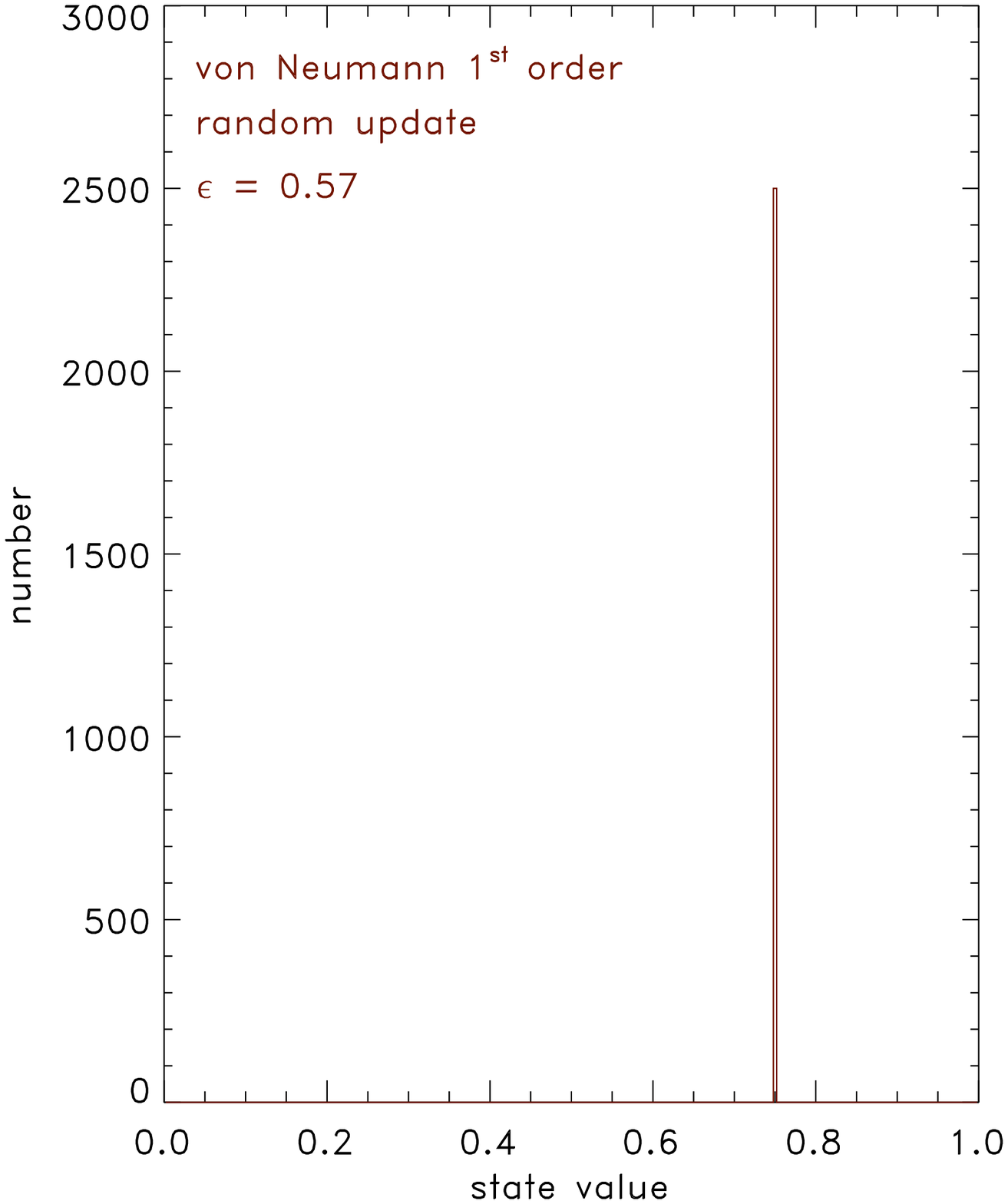,scale=0.45}  
\begin{quote}
{\footnotesize Figure 4: State histograms for random asynchronous updating of CMLs with a von Neumann
neighborhood of order 1 for coupling strengths (a) $\epsilon = 0.47$ (left) and (b) $\epsilon = 0.57$ (right). 
The control parameter of the logistic map is set at $r=4$, and the number of iterations is 10000.}    
\end{quote}
\bigskip
\epsfig{figure=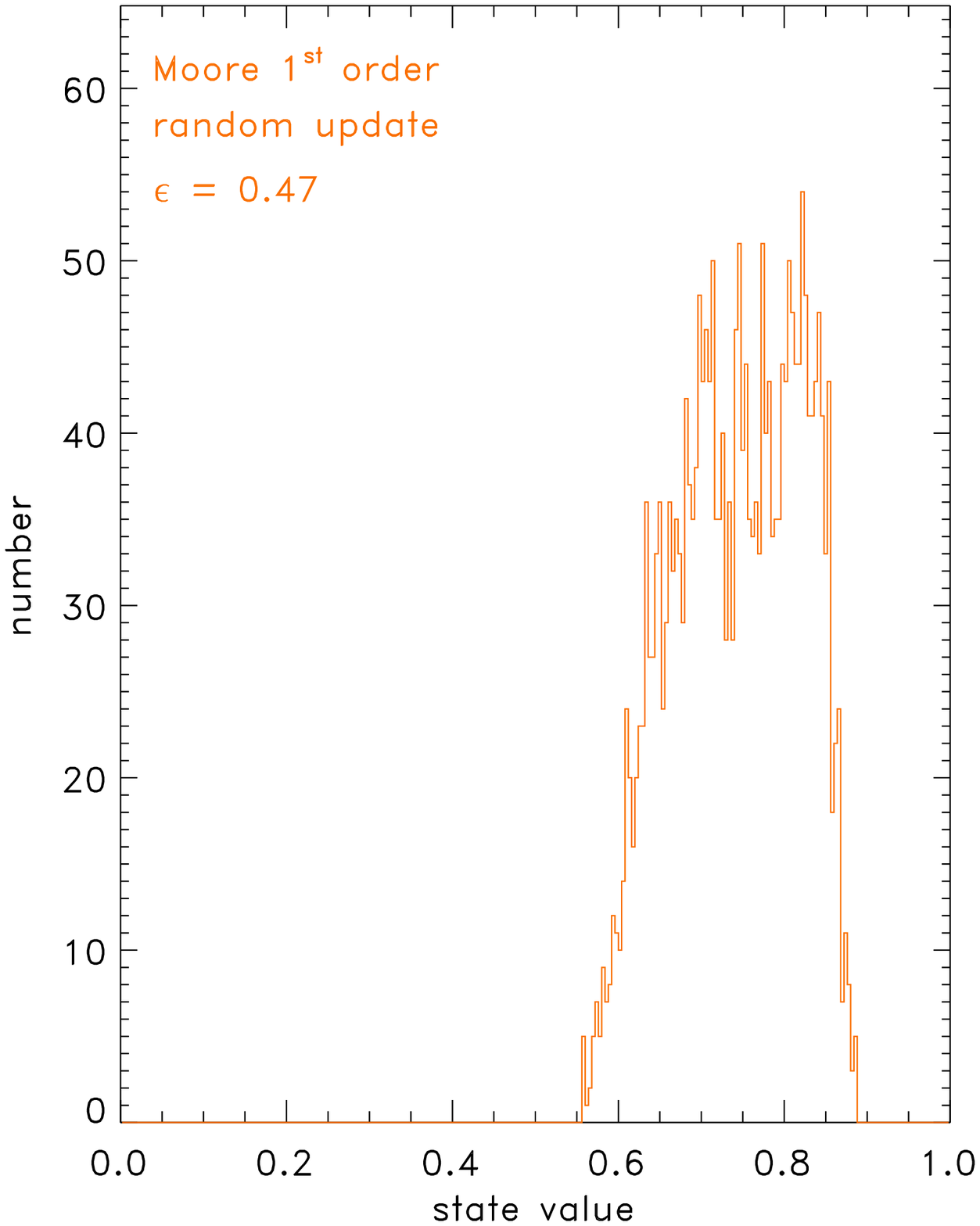,scale=0.45}  
\epsfig{figure=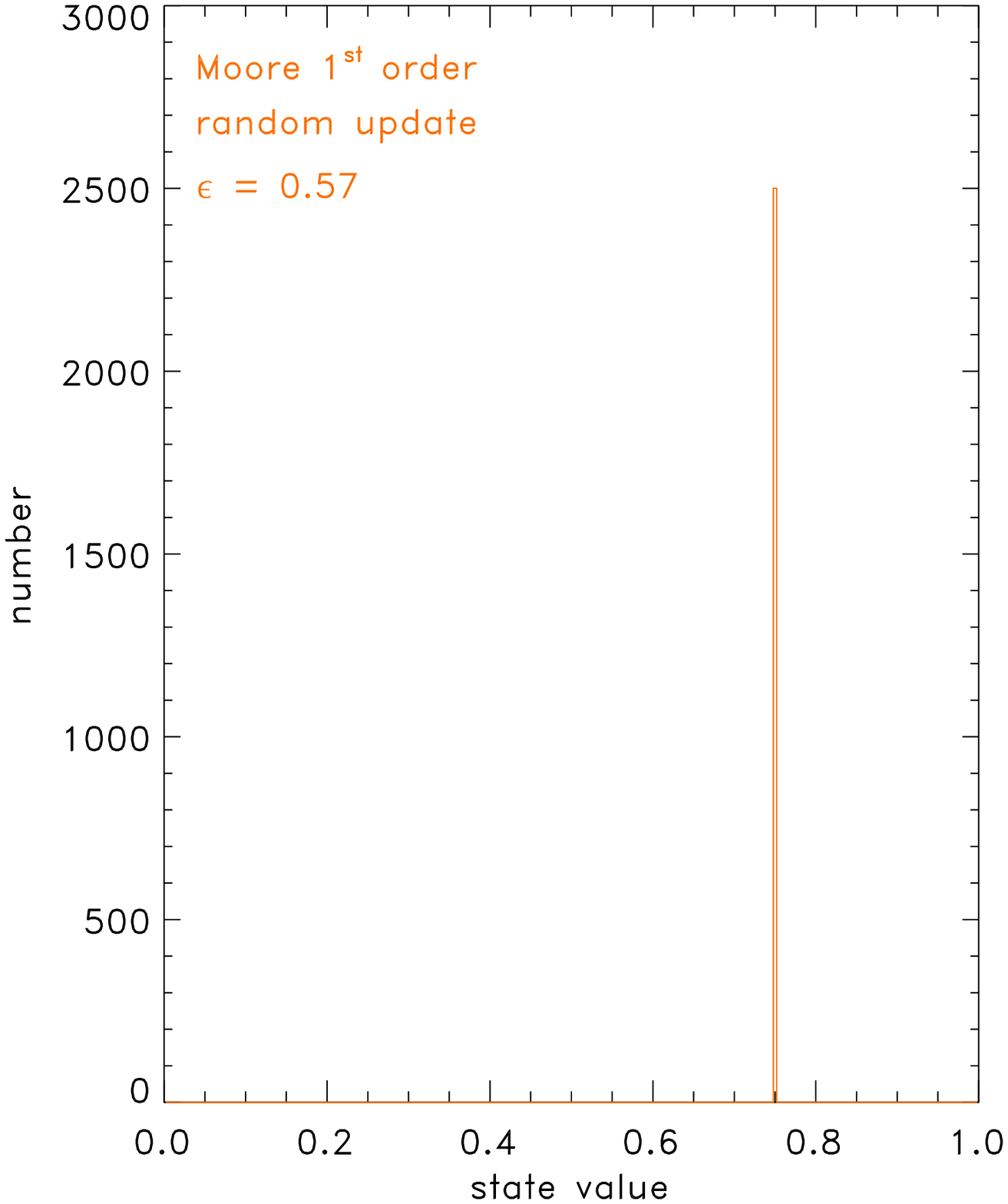,scale=0.45}  
\begin{quote}
{\footnotesize Figure 5: State histograms for random asynchronous updating of CMLs with a 
Moore neighborhood of order 1 for coupling strengths (a) $\epsilon = 0.47$ (left) and (b) $\epsilon = 0.57$
(right). The control parameter of the logistic map is set at $r=4$, and the number of iterations is 10000.}    
\end{quote}
\end{figure}

For {\it random} asynchronous updates, Figs.~4 and 5 show two histograms 
for the distribution of state values after 10000 iteration steps for a von Neumann 
neighborhood of order 1 and a Moore neighborhood of order 1, for coupling strengths 
$\epsilon = 0.47$ (Figs.~4a, 5a) and $\epsilon = 0.57$
(Figs.~4b, 5b). Figs.~4a and 5a show the overall state distribution
of the CML. (For a linear update, using the linear 
sequence of cells as it is given by the structure of the (2D) lattice,
there are two peaks right and left of 
the unstable fixed point at 0.75, indicating oscillatory behavior.) 
By contrast, Figs.~4b and 5b show that the entire CML is stabilized  
at the unstable fixed point at 0.75 at each cell for both types of neighborhood.

\begin{figure}[h]
\begin{center}
\vskip 0.2cm
\epsfig{figure=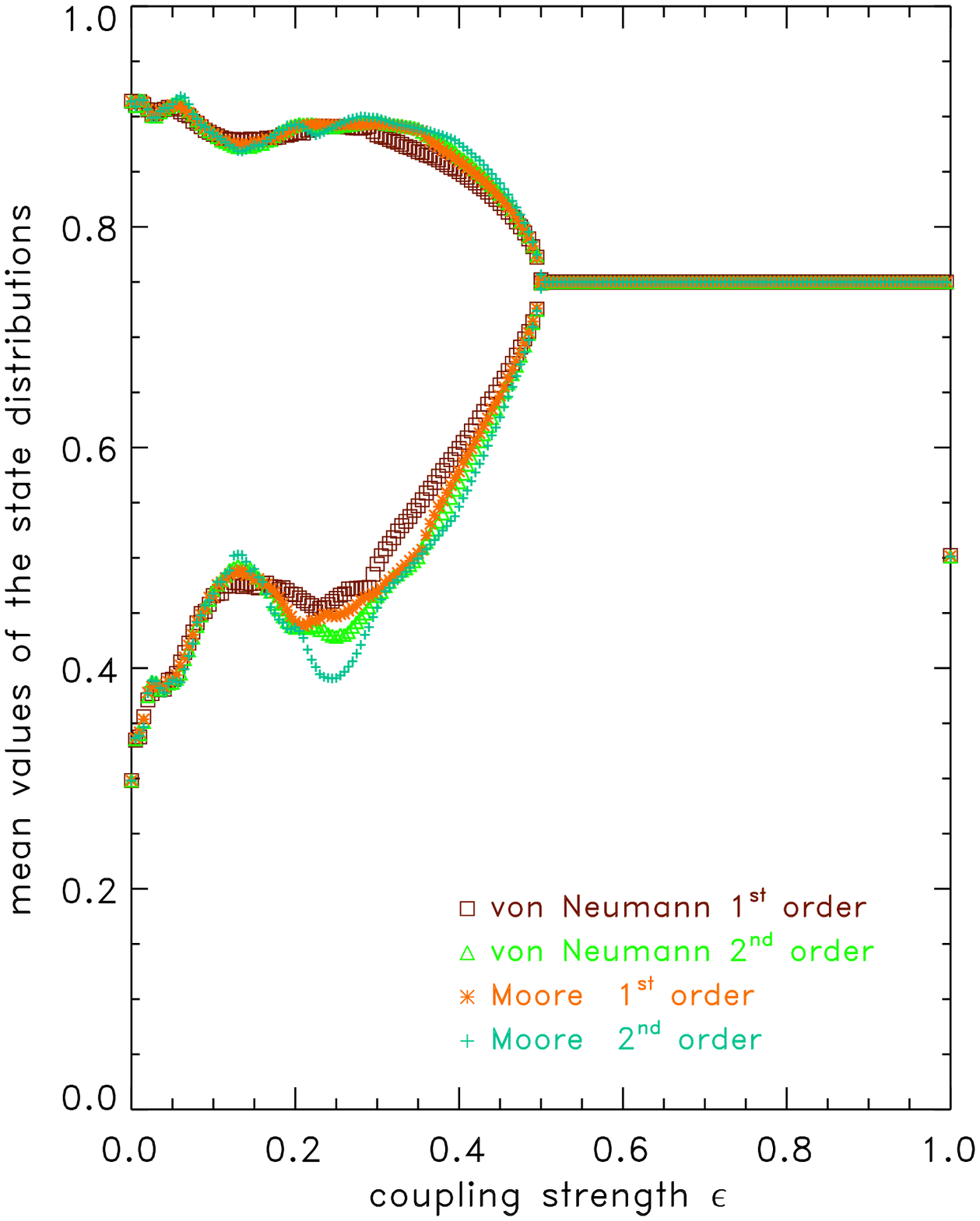,scale=0.6}
\end{center}
\begin{quote} {\footnotesize Figure 6: Stability diagram for CMLs with random asynchronous updating for different 
types of neighborhoods. Mean values of the state distribution right and left of the unstable
fixed point at 0.75, averaged over ten sets of random initial conditions, are plotted
versus the coupling strength $\epsilon$. The control parameter of the logistic map is
set at $r=4$, and the number of iterations is 10000.}    
\end{quote}
\end{figure}

Figure 6 shows the stability diagram for random asynchronous updating. Stabilization 
at the unstable fixed point sets in precisely
at $\epsilon = 0.5$, beyond which all cells stay at the  value 0.75 of the unstable fixed
point. (At $\epsilon=1.0$, this changes to 0.5, the mean value of the 
distribution of initial state values; compare Fig.~3.) 
As for synchronous updating, this behavior occurs for both homogeneous and
inhomogeneous perturbations.
Other than for synchronous updating, the onset of 
stabilization is independent of the chosen neighborhood. The behavior of the stability
curves for $\epsilon < 0.5$ shows small variations for different neighborhood types.

It should be noted here that a {\it linear} asynchronous updating procedure, 
leads to the same stability diagram  
as random asynchronous updating. In this respect, linear updating can be considered as a 
special case of random updating.      

This is not the case for {\it value-dependent} asynchronous updating, where the updating 
sequence is determined by the distribution of state values in the preceding time step.
As an example, this is demonstrated for the case in which the updating sequence corresponds 
to the sequence of decreasing state values. i.e.~ the site with the highest value is updated
first etc. In Fig.~7, histograms for the distribution of state values after 10000 iteration steps
for such a value-dependent asynchronous updating are presented for coupling strength 
$\epsilon = 0.47$, (a) for a von Neumann neighborhood of order 1 and (b) for a Moore 
neighborhood of order 1. For $\epsilon = 0.57$ (and other values of $\epsilon > 0.5$)
the distribution of state values looks 
precisely as for random asynchronous updating in Figs.~4b and 5b. 

\renewcommand{\baselinestretch}{0.85}
\begin{figure}[h]
\epsfig{figure=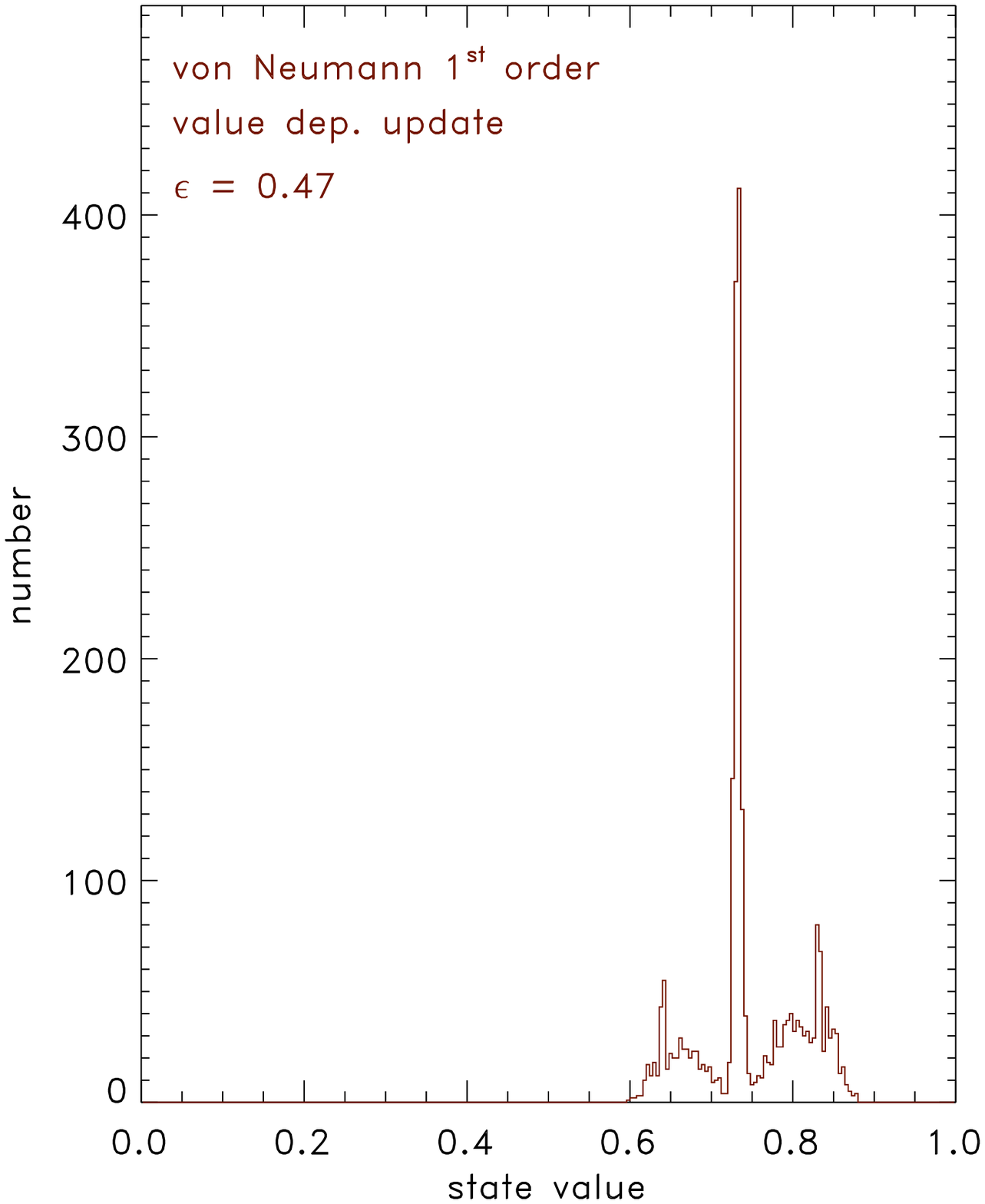,scale=0.45}  
\epsfig{figure=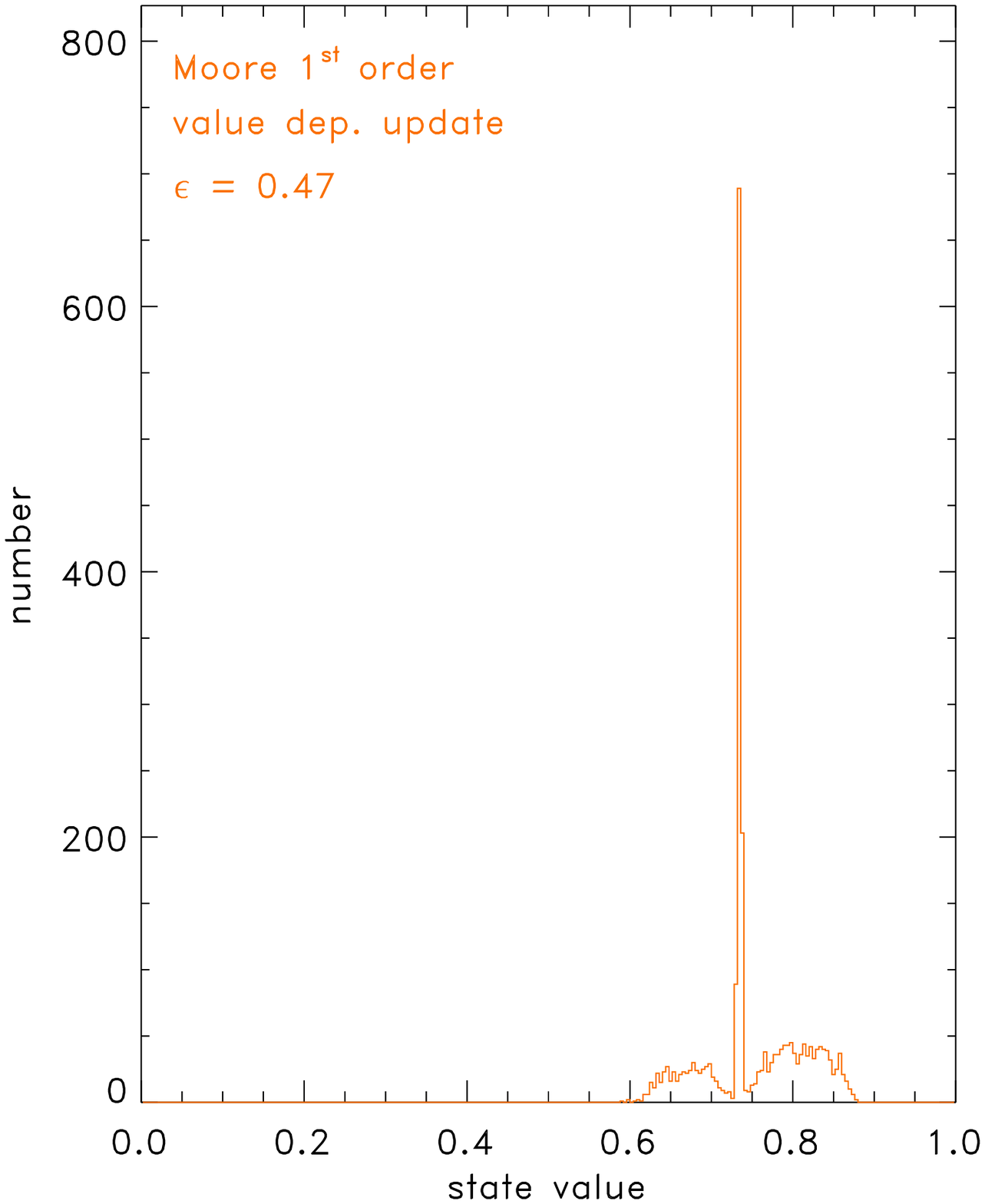,scale=0.45}  
\begin{quote}
{\footnotesize Figure 7: State histograms for value-dependent asynchronous updating of CMLs with (a)
von Neumann neighborhood (left) and (b) 
Moore neighborhood (right) of order 1 for coupling strength $\epsilon = 0.47$. The state histogram
for $\epsilon = 0.57$ is identical with that of Figs.~4b and 5b. 
The control parameter of the logistic map is set at $r=4$, and the number of iterations is 10000.}    
\end{quote}
\end{figure}

Fig.~8 shows the stability diagram for value-dependent asynchronous updating. Its overall
appearance is similar to Fig.~6. However, the lower branch shows a
slight difference in the range $0.4 < \epsilon < 0.5$. 
It reflects the fact that mean values of the state distribution
for states smaller than $x_c$, the value of the unstable fixed point, tend to approach that value
faster than those greater than $x_c$. Another difference is that the mean value
for $\epsilon =1$ is approximately 0.3 rather than 0.5. This reflects the fact that, 
as a consequence of the
value dependent updating, higher values in the map are more often reduced than small values 
are increased, thus leading to a lowered mean for perfect coupling.  

\begin{figure}[h]
\begin{center}
\vskip 0.2cm
\epsfig{figure=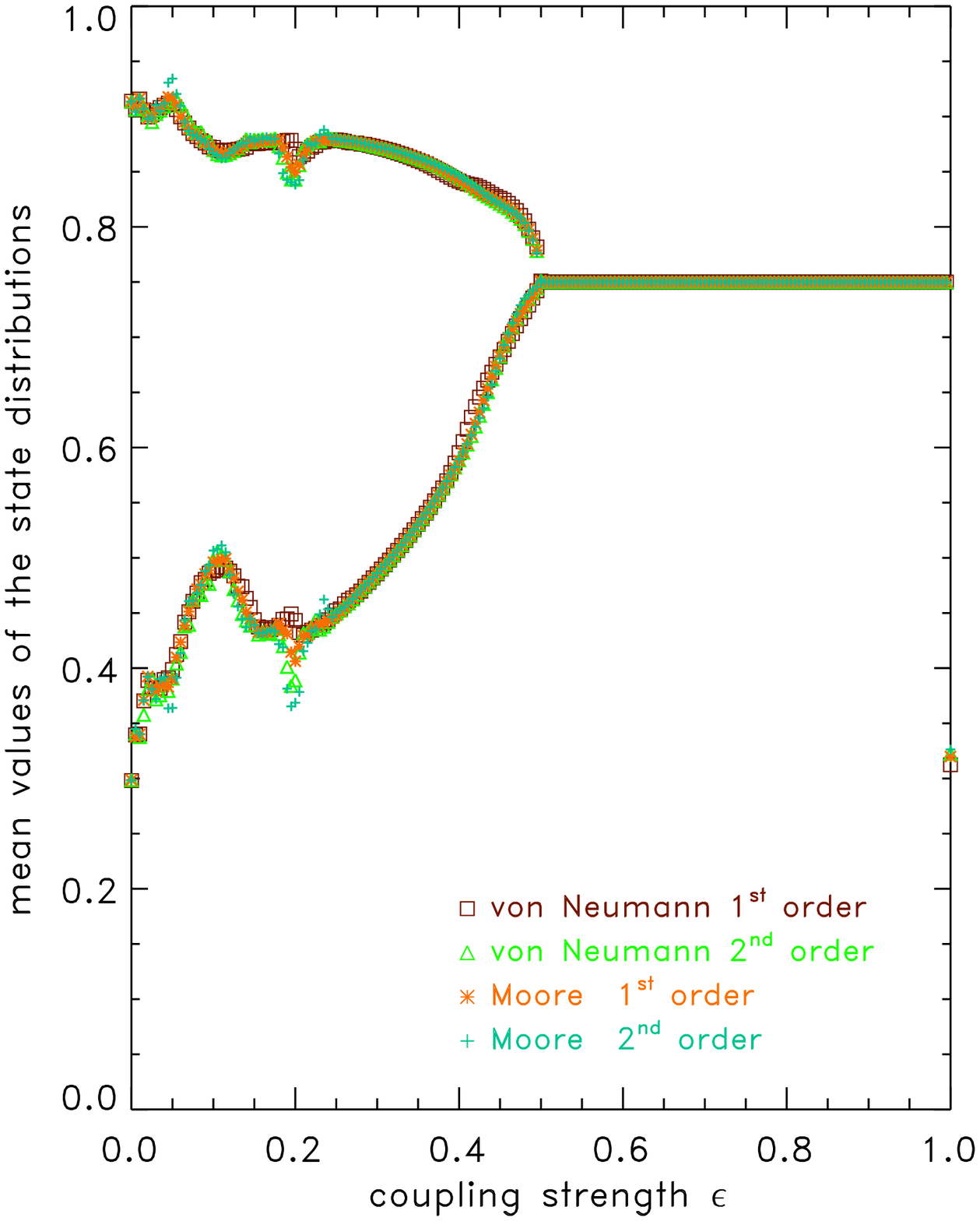,scale=0.6}
\end{center}
\begin{quote} {\footnotesize Figure 8: Stability diagram for CMLs with value-dependent
asynchronous updating for different 
types of neighborhoods. Mean values of the state distribution right and left of the unstable
fixed point at 0.75, averaged over ten sets of random initial conditions, are plotted
versus the coupling strength $\epsilon$. The control parameter of the logistic map is
set at $r=4$, and the number of iterations is 10000.}    
\end{quote}
\end{figure}

\section{Analytical Conditions for Stabilization}  

To some extent it is possible to understand the stabilization behavior of local unstable 
behavior of CMLs analytically. In the first term at the rhs of Eq.~(1), one may consider
$r_{\rm eff} = (1-\epsilon)\ r$ as an ``effective'' control parameter including the influence
of coupling among cells.
Then, the maximum of the resulting ``effective'' logistic map is lowered from $r/4$ to
$r_{\rm eff}/4$, thus leading to a vertically squeezed version of the original map. For this 
squeezed map, the critical point is ${r_{\rm eff}-1}\over r_{\rm eff}$, which is different from 
the original critical point unless $\epsilon = 0$. 

The original critical point can be re-established, if the squeezed map is shifted upward 
until it intersects the line $x_{n+1} = x_n$ at the critical point $x_c = {{r-1}\over r}$. Now, 
the absolute value of the derivative of the squeezed-and-shifted map at $x_c$ is clearly smaller
than that of the original map. In particular, its absolute value is smaller than 1 if $\epsilon$ is large 
enough, such that the fixed point becomes stable. Figure 9 illustrates 
this argument. 

\begin{figure}[h]
\begin{center}
\unitlength 0.7mm
\linethickness{0.4pt}
\begin{picture}(100.00,100.00)
\bezier{5000}(00.00,00.00)(50.00,200.00)(100.00,00.00)
\qbezier(00.00,00.00)(50.00,80.00)(100.00,00.00)
\qbezier(00.00,45.00)(50.00,125.00)(100.00,45.00)
\put(0,0){\framebox(100,100)}
\put(0.00,0.0){\line(1,1){100.00}}
\put(75.00,30.00){\line(0,1){45.00}}
\put(73.00,45.00){\makebox(0,0)[cc]{$d$}}
\put(70.00,15.00){\makebox(0,0)[cc]{\footnotesize squeezed}}
\put(70.00,10.00){\makebox(0,0)[cc]{\footnotesize map}}
\put(50.00,75.00){\makebox(0,0)[cc]{\footnotesize squeezed-and}}
\put(50.00,70.00){\makebox(0,0)[cc]{\footnotesize -shifted map}}
\put(15.00,90.00){\makebox(0,0)[cc]{\footnotesize original}}
\put(15.00,85.00){\makebox(0,0)[cc]{\footnotesize map}}
\put(75.00,75.00){\circle*{2}}
\put(37.00,37.00){\circle*{2}}
\put(75.00,30.00){\circle{2}}
\put(82.00,12.50){\vector(1,0){5}}
\put(25.00,87.50){\vector(1,0){5}}
\put(50.00,78.00){\vector(0,1){5}}
\end{picture}
\end{center}
\begin{quote}
{\footnotesize Figure 9: Schematic illustration of the squeeze-and-shift procedure 
explained in the text. The unstable fixed point of the original logistic map and
the stable fixed point of the squeezed map are represented as filled dots; the empty
dot indicates the value of the squeezed map at the unstable fixed point of the
original map. In order to match the unstable fixed point of the original map, the
squeezed map has to be shifted by the distance $d$.}
\end{quote}
\end{figure}

The unstable fixed point of the logistic map is stabilized if the distance $d$ between the
original logistic map at $x_{c}$ and the squeezed map at the same value is  
compensated by the shift due to the neighborhood term in Eq.~(1). 
The distance $d$ (cf.~Fig.~9) is given by:
\begin{eqnarray}
d &=& x_c - f_{\rm sqeezed}(x_c) \\
&=& x_c - rx_c (1-\epsilon)(1-x_c) 
\end{eqnarray} 
For stabilization, this distance must be equal to the second term in Eq.~(1):
\begin{equation}
d = {\epsilon\over N} \sum_{k=1}^N x_k 
\end{equation}
Combining Eqs.~(3) and (4) yields: 
\begin{eqnarray}
{1\over N} \sum_{k=1}^N x_k &=& {1\over\epsilon} x_c \left[1-r(1-\epsilon)(1-x_c)\right] \\
&=&  x_c
\end{eqnarray}
since some arithmetic shows that: 
\begin{equation}
\left[1-r(1-\epsilon)(1-x_c)\right] = \epsilon
\end{equation}

Thus, a first necessary condition for 
stabilization is that the mean of the state values in all neighborhood cells 
is exactly the critical value $x_c$. 
The distance $d = \epsilon x_c$ itself is independent of the size $N$ of the neighborhood.

The values of $\epsilon$ and $r$ which provide stabilization cannot be determined from the 
derived condition. This means that the quantitative dependence of the stabilization on
these parameters cannot be inferred from Eq.~(6). 
Of course, we know that $r>3$ is required for an unstable fixed point at $r-1\over r$. 
Moreover, it is evident from the numerical results presented in Sec.~2 that $\epsilon$ 
plays a crucial role for stabilization. 

The value of $\epsilon$ required to make the derivative of the squeezed-and-shifted 
map equal to -1 at the critical point $x_c$, thus leading to stabilization, can be obtained 
from the derivative of the squeezed map at $x_c$: 
\begin{equation}
f'(x_c) = r_{\rm eff} (1-2x_c)
\end{equation}
Setting $f'(x_c) = -1$ yields
\begin{equation}
\epsilon_{lb} = 1 - {1\over r-2} 
\end{equation}
as the lower bound for the coupling strength $\epsilon_c$ of stabilization onset. 
This is a second
necessary condition for the global stabilization of an unstable local fixed point. 
It expresses the fact that the coupling among lattice cells leads to stable behavior
as reflected by the squeezed map in Fig.~9, so that the unstable fixed point at individual cells
becomes inefficacious. This explains why the stabilization does not depend on whether the
perturbation is homogeneous or inhomogeneous, as observed in Secs.~2.1 and 2.2.  

For $r=4$, it follows 
that $\epsilon_c \ge 0.5$. As can be seen
in Figs.~6 and 8, the onset of stabilization for asynchronous updating at $r=4$ is precisely at
$\epsilon_{lb} = \epsilon_c = 0.5$, irrespective of the type of neighborhood chosen and irrespective of the
way in which the updating sequence is determined. For $r=3$, $\epsilon_{lb} = \epsilon_c = 0$ since the 
fixed point at $r-1\over r$ becomes a stable fixed point.   
Figure 10 shows the functional dependence of $\epsilon_{lb}$ on $r$ according to Eq.~(9) 
together with some values of $\epsilon_c$ obtained from numerical runs for random 
asynchronous updating with different values of $r$. 
The numerical results $\epsilon_c$ match their lower bound $\epsilon_{lb}$ perfectly well.   

\begin{figure}[h]
\begin{center}
\vskip 0.2cm
\epsfig{figure=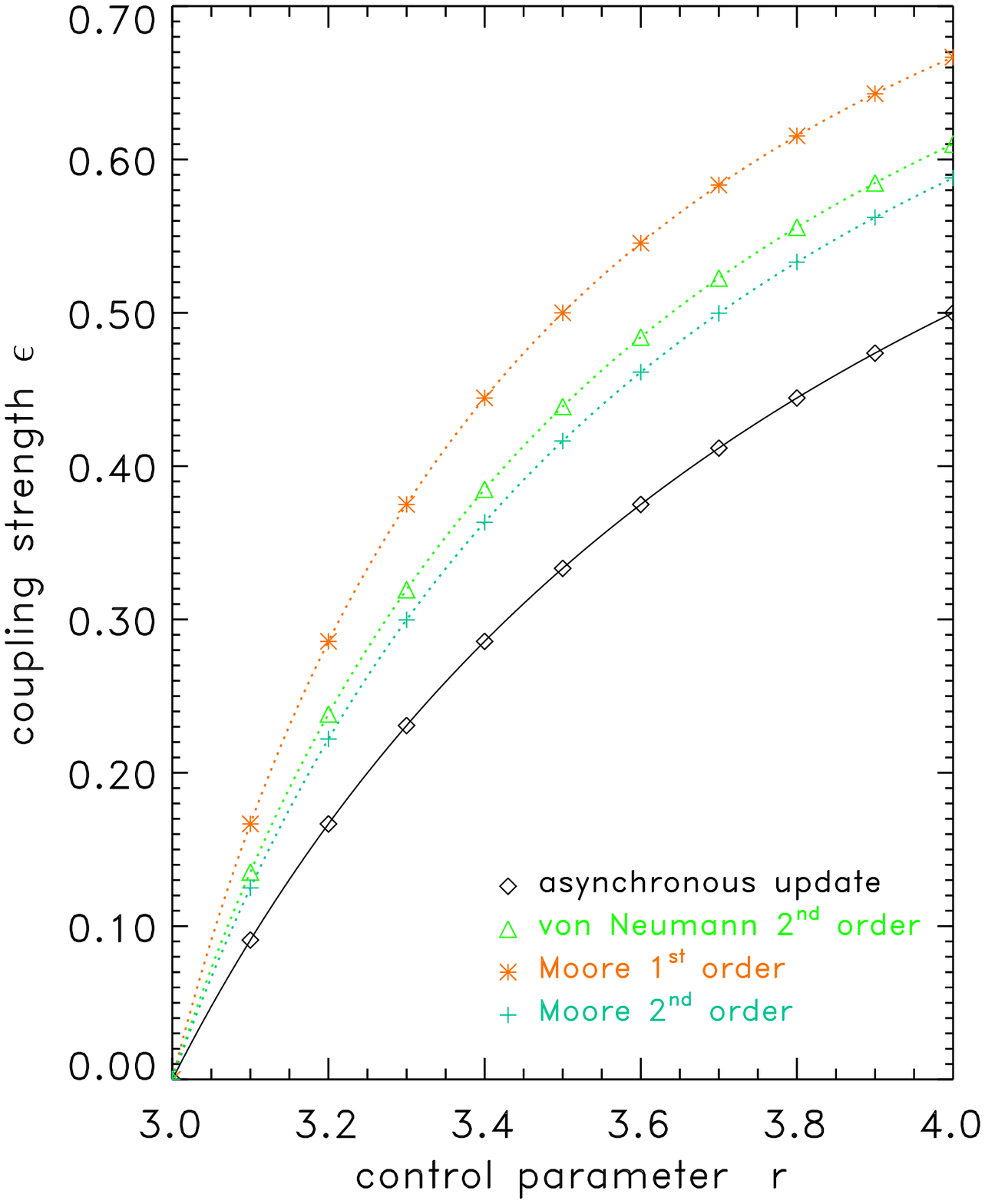,scale=0.5}
\end{center}
\begin{quote} {\footnotesize Figure 10: Coupling strength for stabilization onset, $\epsilon_c$, 
and its lower bound, $\epsilon_{lb}$, as a function of $r$, $3 \le r \le 4$. 
The solid curve for $\epsilon_{lb}$ is obtained from Eq.~(9). The individual values 
along the curve are numerical results of $\epsilon_c = \epsilon_{lb}$ for random asynchronous 
updating. Values above the theoretical curve represent $\epsilon_c > \epsilon_{lb}$ 
for synchronous updating with different neighborhoods. The dotted curves represent 
correspondingly rescaled versions of Eq.~(9).}    
\end{quote}
\end{figure}

For synchronous updating (cf.~Fig.~3), the onset of stabilization (except for a von Neumann 
neighborhood of order 1) satisfies the lower bound according to Eq.~(9) as well. However, the value
$\epsilon_c$ of stabilization onset is strictly greater than $\epsilon_{lb}$ and 
depends additionally on the neighborhood. This can also be seen in Fig.~10 where $\epsilon_c$ 
is plotted for different kinds of neighborhood. In order to describe
the corresponding behavior analytically (dotted lines in Fig.~10), 
$r$ in Eq.~(9) has to be properly rescaled in a way 
which depends on the size and topology of the neighborhood.                 


\section{Summary and Perspectives} 

The behavior of two-dimensional lattices of coupled logistic maps with different 
coupling $\epsilon$, different types of neighborhoods and different updating rules 
has been studied at and around
the unstable fixed point of individual maps. Moore and von Neumann 
neighborhoods up to second order as well as synchronous and asynchronous
updating rules have been considered. It has been found that locally
unstable behavior of individual maps is stabilized due to the influence of
global lattice interactions for particular parameter ranges. The key point
of this stabilization is that it operates inherently, i.e.~without any need for 
external adjustment or control.       

For synchronous updating, where all sites of the 
lattice are updated simultaneously, stabilization has been observed for
all types of neighborhood with the exception of a 
von Neumann neighborhood of first order.  The onset value $\epsilon_c$ for  
stabilization depends on the type of neighborhood and, of course, on the 
control parameter $r$ of the logistic map. 

For asynchronous updating the sites of the lattice are updated sequentially.
This can be straightforwardly implemented in terms of a randomized
sequence (or, somewhat more articifially, linearly with respect to the structure
of the lattice). Another option is a value-dependent updating procedure,
where the updating sequence is determined by the sequence of ordered 
state values. In both cases, stabilization is always achieved. The coupling
strength $\epsilon_c$ for its onset depends on $r$, but not on the type of 
neighborhood.      

Two conditions for stabilization have been derived analytically. The first one
states that the sum of the values in all $N$ neighborhood cells must be
equal to $N$ times the critical value of the map in order to achieve stabilization. 
The second condition provides a lower bound for the coupling strength $\epsilon_c$ 
required for stabilization onset. This lower bound is precisely the onset
value itself in case of asynchronous updating, but not in case of 
synchronous updating.       

More complicated behavior of coupled map lattices 
can easily be generated by more complicated
lattice features. For instance, first pilot studies combining random and 
value-dependent asynchronous 
updating have shown situations in which an onset of stabilization at some $\epsilon$
is observed, but the lattice destabilizes again with increasing coupling strength.    

The reported results have been obtained for homogeneous and time-independent
coupling. Relaxing these conditions leads to plastic networks as 
recently investigated by Ito \& Kaneko [2000, 2002]. Another important novel direction
of research concerns so-called scale-free networks (Albert \& Barabasi [2002]), where
the number of edges, which a node in the network has, is not Poisson distributed
(as in a standard random graph) but shows power-law behavior.    

\section{References}

\begin{description}

\item Albert, R., \& Barabasi, A.-L.~[2002]: ``Statistical mechanics of complex networks'',
{\it Reviews of Modern Physics} {\bf 74}, 47--97.

\item Atay, F., Jost, J., \& Wende, A.~[2004]: ``Delays, connection topology,
and synchronization of coupled chaotic maps'', lanl preprints cond-mat/0312177.

\item Atmanspacher, H.~[1992]: ``Categoreal and acategoreal representation of 
knowledge'', {\it Cognitive Systems} {\bf 3}, 259--288.

\item Atmanspacher, H., \& Wiedenmann, G.~[1999]: ``Some basic problems with complex 
systems'', in {\it Large Scale Systems: Theory and Applications}, ed.~by 
N.T.~Koussoulas and P.~Groumpos, Elsevier, Amsterdam, pp.~1059--1066.     

\item Hopfield, J.J.~[1982]: ``Neural networks and physical systems with emergent
collective computational abilities'', {\it Proceedings of the National Academy of Sciences 
of the USA} {\bf 79}, 2554--2558.  

\item Ito, J., \& Kaneko, K.~[2000]: ``Self-organized hierarchical structure
in a plastic network of chaotic units'', {\it Neural Networks} {\bf 13},
275--281.

\item Ito, J., \& Kaneko, K.~[2002]: ``Spontaneous structure formation in a
network of chaotic units with variable connection strength'', {\it Physical
Review Letters} {\bf 88}, 028701.  

\item Kaneko, K., ed.~[1993]: {\it Theory and Applications of Coupled Map 
Lattices}, Wiley, New York.

\item Kaneko, K., \& Tsuda, I.~[2000]: {\it Complex Systems: Chaos and Beyond},
Springer, Berlin.

\item Kornmeier, J., Bach, M., \& Atmanspacher, H.~[2004]: ``Correlates of perceptive 
instabilities in event-related potentials'', {\it International Journal of Bifurcations 
and Chaos} {\bf 14}, 727--736.

\item Kuhn, A., Aertsen, A., \& Rotter, S.~[2004]: ``Neuronal integration of synaptic 
input in the fluctuation-driven regime'', {\it Journal of Neuroscience} {\bf 24}, 2345--2356.

\item Li, C., Li, S., Liao, X., \& Yu, J.~[2004]: ``Synchronization in coupled
map lattices with small-world delayed interactions'', 
{\it Physica A} {\bf 335}, 365--370.

\item Lumer, E.D., \& Nicolis, G.~[1994]: ``Synchronous versus asynchronous dynamics
in spatially distributed systems'', {\it Physica D} {\bf 71}, 440--452.  

\item Mackey, M., \& Milton, J.~[1995]: ``Asymptotic stability of densities in coupled 
map lattices'', {\it Physica D} {\bf 80}, 1--17.

\item Marcq, P., Chat\'e, H., \& Manneville, P.~[1997]: ``Universality in Ising-like
phase transitions of lattices of coupled chaotic maps'', {\it Physical Review E}
{\bf 55}, 2606--2627. 

\item Masoller, C., Marti, A.C., \& Zanette, D.H.~[2003]: ``Synchronization in an
array of globally coupled maps with delayed interactions'', {\it Physica A}
{\bf 325}, 186--191.

\item Mehta, M., \& Sinha, S.~[2000]: ``Asynchronous updating of coupled maps leads
to synchronization'', {\it CHAOS} {\bf 10}, 350--358. 

\item Ott, E., Grebogi, C., \& Yorke, J.A.~[1990]: ``Controlling chaos'',
{\it Physical Review Letters} {\bf 64}, 1196--1199.

\item Rolf, J., Bohr, T., \& Jensen, M.H.~[1998]: ``Directed percolation universality
in asynchronous evolution of spatiotemporal intermittency'', {\it Physical Review E}
{\bf 57}, R2503--R2506. 


\item Turing, A.~[1952]: ``The chemical basis of morphogenesis'', {\it Transactions
of the Royal Society London, Series B} {\bf 237}, 37--72.

\end{description}
\end{document}